\UseRawInputEncoding
\documentclass[english]{cccconf}
\usepackage[comma,numbers,square,sort&compress]{natbib}
\usepackage{epstopdf}

\usepackage{amsmath} 
\usepackage{amssymb}
\usepackage{graphicx}
\usepackage{svg}
\begin{document}

\title{Co-Optimization of Adaptive Cruise Control and\\ Hybrid Electric Vehicle Energy Management\\ via Model Predictive Mixed Integer Control}

\author{Qitao Li\aref{amss},
        Changfu Gong\aref{amss},
        Yuan Lin\aref{amss}}



\affiliation[amss]{Shien-Ming Wu School of Intelligent Engineering, South China University of Technology, Guangzhou 511442, P.~R.~China
        \email{yuanlin@scut.edu.cn}}

\maketitle

\begin{abstract}
In this paper, a model predictive mixed integer control method for BYD Qin Plus DM-i (Dual Model intelligent) plug-in hybrid electric vehicle (PHEV) is proposed for co-optimization to reduce fuel consumption during car following. First, the adaptive cruise control (ACC) model for energy-saving driving is established. Then, a control-oriented energy management strategy (EMS) model considering the clutch engagement and disengagement is constructed. Finally, the co-optimization structure by integrating ACC model and EMS model is created and is converted to the mixed integer nonlinear programming (MINLP). The results show that this modeling method can be applied to EMS based on the model predictive control (MPC) framework and verify  that co-optimization can  achieve a 5.1$\%$ reduction in fuel consumption compared to sequential optimization with the guarantee of ACC performance.
\end{abstract}


\footnotetext{This work is supported by Guangzhou basic and applied basic research project under Grant 2023A04J1688.}

\section{Introduction}

In 2021, China$'$s permeability of new energy vehicle market increase to 13.4$\%$ rapidly\cite{Zhang2023}. As one of the most popular new energy vehicles, PHEVs are focused by current paper in ACC condition\cite{Yang2021}. 

ACC is the most widely used advanced driving assistance system (ADAS) and the common velocity planning method. Through smoothing the velocity profile, ACC reduces the fuel consumption, improves vehicle comfortability and safety\cite{Chen2018}. To further reduce fuel consumption, Li et al. presented an ecological adaptive cruise controller (ECO-ACC) with a two-level framework for a PHEV, which could maintain a proper distance from the preceding vehicle\cite{G.Li2018}.

The main reasons for PHEVs can be energy-saving are regenerative braking, engine stopping and operating in high-efficiency area. PHEVs combine the advantages of electric vehicles (EVs) and hybrid electric vehicles (HEVs), which have the capability of pure electric driving\cite{Lin2022}. Li et al. illustrated the outstanding energy saving of DM-i by simulation and experiment\cite{Li2021}. BYD Qin Plus DM-i has the forced charge sustaining mode, which allows battery state of charge (SOC) to be maintained around the set value in order to ensure the acceleration performance and to cope with situations where external power grid is not available. Currently, there is lack of literature modeling the control-oriented models for BYD Qin Plus DM-i using optimization-based method .

The purpose of EMS for HEVs or PHEVs is to minimize energy consumption by deciding the speed and torque between the engine and the motor. There are three main categories to EMS: rule-based, optimization-based, and learning-based. Rule-based methods specify the torque of the engine and motor according to SOC and torque demand\cite{Banvait2009}. Optimization-based energy management includes global optimization strategies such as dynamic programming (DP), Pontryagin's minimum principle (PMP), and the local optimization strategies such as MPC\cite{Robuschi2017}. Reinforcement learning (RL) can achieve fuel consumption close to DP\cite{Lian2020}.

The optimization structure composed of velocity planning and EMS can be classified into two categories: sequential optimization and co-optimization. Sequential optimization optimizes velocity planning and energy management separately. The optimal velocity or acceleration command is obtained by the upper controller, which is input into the lower EMS controller to minimize energy consumption. Due to the real-time performance and the simplicity of modeling and solving, large numbers of studies focus on it. The upper and lower controllers could be applied to the same or different solutions, such as convex optimization, alternating direction method of multipliers (ADMM)\cite{Liu2022}, DP and deep reinforcement learning (DRL)\cite{Tang2021}. These solutions can be connected in series as different sequential optimizations, which lead to different fuel consumption and calculation time. However, the upper controller gives the desire speed or acceleration in advance, which limits the EMS optimization space.

Co-optimization optimizes velocity planning and energy management as an integration. Due to a wider feasible region, the better optimal solution can be found to achieve lower fuel consumption in co-optimization. Co-optimization achieves reduction in total energy consumption compared to sequential optimization\cite{Li2017}. In some traffic scenarios, such as velocity limit\cite{Liu2022} or traffic light constraints\cite{Liu2021}, the global optimal results of co-optimization are solved by DP, which can be used as a benchmark for comparison with other methods. However, ACC requires short sampling time and accurate acceleration command. The high-dimensional and high-precision grids make DP to solve the co-optimization of ACC and EMS difficultly. In other words, it is a challenge to solve the global optimal results of co-optimization based on ACC and EMS in drive cycle. Using MPC framework to solve the co-optimization problem is one of the most common methods\cite{Chen2020, Chen2021, Chen2022}. MPC can effectively deal with multi-objective optimization problems with constraints, which has been widely used in ACC and EMS. The short prediction horizon can reduce the computation burden and enable the co-optimization problem to be solved.  Based on this, we use the MPC framework to achieve both co-optimization and sequential optimization.

The main contributions and innovative points of this paper include three aspects:

\begin{itemize}
  \item A modeling method for BYD Qin Plus DM-i is proposed, which includes the engine, motor, generator model, etc.
  \item Considering the engagement and disengagement of clutch, the EMS problem is converted to MINLP.
  \item By comparing with sequential optimization, the comfort, energy saving and computational burden of co-optimization are tested.
\end{itemize}

The remainder of this paper is organized as follows: Section 2 introduces the modeling of ACC model. Section 3 details the DM-i system. Section 4 presents the structures of sequential optimization and co-optimization based on MPC. Section 5 details the simulation and analyses the optimization result. Section 6 concludes the paper and prospects future work.

\section{Adaptive Cruise Control Model}

ACC model can be described as\cite{Li2019}
\begin{align}\label{eq:acc}
\quad\dot{s} &=v_r\\
\dot{v_r}& =a_p-a_h
\end{align}

\noindent where $s$ is the actual distance between the preceding vehicle and the host vehicle
, $v_r=v_p-v_h$ is the relative velocity, $v_h$ and $v_p$ are the velocity of the host vehicle and the preceding vehicle, $a_h$ and $a_p$ are the acceleration of the host vehicle and the preceding vehicle.

The optimal distance area is written as
\begin{align}\label{eq:acc}
\quad s_{opt,min} &=5.2+0.7v_h+0.0705v_h^2\\
\quad s_{opt,max} &=6.8+0.8v_h+0.0745v_h^2
\end{align}

\noindent where $s_{opt,min}$ and $s_{opt,max}$ are the minimum and the maximum of optimal distance\cite{Li2017}. The optimal distance area gives the host vehicle a more flexible time headway to avoid aggressive following, which helps to reduce fuel consumption. Maintaining the small time headway could lead the host vehicle to follow the preceding vehicle in a more aggressive way, which causes drastic acceleration and more fuel consumption\cite{Prakash2016}.

The distance error is defined as
\begin{align}\label{eq:e}
e=
\begin{cases}
s_{opt,min}-s,& \text{ if }  s\textless s_{opt,min} \\
0, &\text{ if }  s_{opt,min}\leq s\leq s_{opt,max} \\
s-s_{opt,max},& \text{ if }  s\textgreater s_{opt,max}
\end{cases}
\end{align}

\noindent When the actual distance $s$ locates in the optimal distance area, the value of $e$ will be zero.

Jerk is defined as the rate of change in $a_h$, which is relative to the driving comfort of the host vehicle.
\begin{align}\label{eq:acc}
\quad j_h &=\Delta a_h
\end{align}

\section{DM-i Model}

\begin{figure}[!htb]
  \centering
  \includegraphics[width=\hsize]{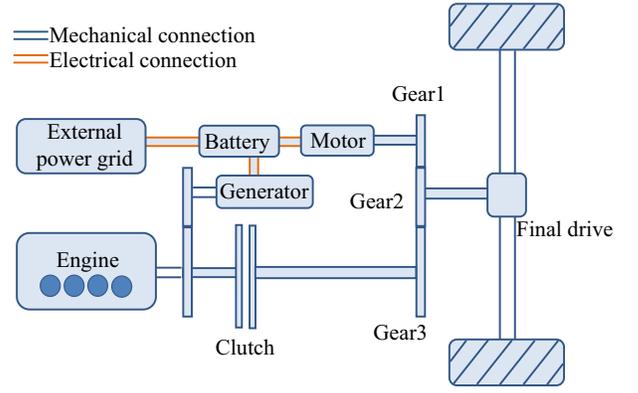}
  \caption{ Powertrain configuration of the BYD Qin Plus DM-i.}
  \label{fig1}
\end{figure}

The simulation model of BYD Qin Plus DM-i is shown in Fig.~\ref{fig1}. This series-parallel PHEV has an engine, a motor, a generator, a battery and a clutch. The vehicle parameters are listed in Table~\ref{tab1} .

\begin{table}[!htb]
  \centering
  \caption{Parameters of BYD Qin Plus DM-i}
  \label{tab1}
  \begin{tabular}{c|c}
    \hhline

    Parameter  & Value \\ \hline
    Vehicle mass $m$ & 1500kg \\ \hline
    Vehicle cross section area $A$ & 2.36m$^2$ \\ \hline
    Wheel radius $r$ & 0.315m \\ \hline
    Rolling resistance coefficient $\mu$ & 0.01 \\ \hline
    Air density $ \rho $  & 1.1985 \\ \hline
    Air drag coefficient $C_d$  & 2.8 \\ \hline

    \hhline
  \end{tabular}
\end{table}

\subsection{Vehicle Dynamics}
According to the longitudinal dynamic model of the vehicle, wheel rolling resistance torque $T_w$ is written as
\begin{align}\label{eq:veh_dyn}
T_w = [ma_h + 0.5 C_d \rho& A v_h^2 + \mu mg cos(\theta) +  mg sin(\theta)] r
\end{align}

\noindent In Eqn.~7 , $m$ is the vehicle mass, $C_d$ is the air drag coefficient, $A$ is the vehicle cross section area, $\mu$ is rolling resistance coefficient, $g$ is the gravity constant, and $\theta$ is the road grade. In this paper, the road grade $\theta$ is considered as 0. There is no rolling resistance when velocity is zero.

The wheel speed is written as 
\begin{align}\label{eq:veh_dyn}
w_{wheel}  = \frac{v_h}{r}
\end{align}
\noindent where $r$ is wheel radius.

\subsection{Engine Model}

The engine speed is written as
\begin{align}\label{eq:engine}
 w_e & = w_{wheel} i_3 i_d i_c+w_e(1-i_c)
\end{align}

\noindent where $w_e$ is the engine speed, $i_3$ is the transmission ratio of gear 3 and gear 2, $i_d$ is gear ratio of the final drive, $i_c$$\in$$\{0,1\}$ is the clutch engagement/disengagement command (0 corresponding to disengagement, 1 to engagement). When the clutch is disengaged, $w_{wheel}$ will not be directly driven by the engine.

The fuel consumption per second $\dot{m}_f$ can be expressed as a piecewise function. The unit of fuel consumption rate $b_e$ is g/(kWh).
When engine torque $T_e$ is larger than 0, a quadratic function of $w_e$ and $T_e$ is used to approximate $\dot{m}_f$.
\begin{align}\label{eq:mfdot}
\dot{m}_f =b_e w_e T_e
\begin{cases}
f_f(w, T_e), \text{ if } T_e > 0\\
0, \text{ if } T_e = 0
\end{cases}
\end{align}

Considering that BYD Qin Plus is a series-parallel PHEV, operation points of engine is unnecessary to distribute in the whole engine map. In order to improve fuel economy, operation points of engine are constrained in high-efficiency area. The engine speed $w_e$ and engine torque $T_e$ are both set to semi-continuous variables shown in Eqns.~11 and 12. The engine could either be off or work in high-efficiency area.
\begin{align}\label{eq:engine}
w_e  &= \text{{$0 \cup {w_{e,min}\leq w_e\leq w_{e,max}}$}}\\
T_e & = \text{{$0 \cup {T_{e,min}\leq T_e\leq T_{e,max}}$}}
\end{align}

\subsection{Motor Model}

The motor speed and motor torque are written as
\begin{align}\label{eq:motor_torque}
w_m&=w_{wheel} i_mi_d\\
T_m = &\frac{T_w  -T_3 i_3 i_d  \eta_t i_c}{i_m i_d  \eta_t }
\end{align}

\noindent where $i_m$ is the transmission ratio of gear 1 and gear 2, $T_3$ is torque transmitted by gear 3, $\eta_t$ is the mechanical efficiency of gears. Eqn.~14 illustrates that when the clutch is disengaged, the gear 3 will not transmit the torque from the engine, and the motor provides the torque for the wheels only.

The motor efficiency is defined as follows.
\begin{align}\label{eq:eta m}
\eta_m =
\begin{cases}
f_1(w_m, |T_m|), \text{ if }  w_{m,min}\leq w_m\textless w_{m，1}  \\
f_2(w_m, |T_m|), \text{ if }  w_{m1}\leq w_m\leq w_{m，2}  \\
f_3(w_m, |T_m|), \text{ if }  w_{m2}\textless w_m\leq w_{m,max}
\end{cases}
\end{align}

\noindent To improve the fitting accuracy, $\eta_m$ is defined as the piecewise function. $|T_m|$ is the absolute value of $T_m$. $f_1$, $f_2$ and $f_3$ are the binary quadratic functions of $w_m$ and $|T_m|$.

\subsection{Generator Model}

The generator speed and generator torque are written as follows.
\begin{align}\label{Tg}
w_g  = \frac{w_e}{i_g}\\
T_g =(T_e-& \frac{{T_3}{i_c}}{\eta_t})i_g \eta_t
\end{align}

In Eqn.~18, the generator efficiency $\eta_g$ is defined as a binary quadratic function of $w_g$ and $|T_g|$.
When the clutch is engaged ($i_c=1$), The engine torque $T_e$ could be distributed to Gear 3 and generator. When the clutch is disengaged ($i_c=0$), the engine torque $T_e$ would be distributed to generator only. Considering the high speed transmitted by the engine to the generator, only a segment of function is needed for fitting.
\begin{align}\label{eq:eta g}
\eta_g =f_g(w_g, |T_g|)
\end{align}

\subsection{Battery Model}

The battery power $P_b$ is composed of motor power and generator power as follows.
\begin{align}\label{eq:pb}
P_b  = &T_m w_m \eta_m^{sgn(-T_m)}-T_g w_g \eta_g
\end{align}

The rate of change in $SOC$ is represented by
\begin{align}\label{eq:soc}
\quad&\dot{SOC} = - \frac{V_{oc}-\sqrt{V_{oc}^2-4R_bP_b}}{2R_bQ_{max}}
\end{align}

\noindent where $Q_{max}$ is the battery capacity, $V_{oc}$ is the open circuit voltage and $R_b$ is the internal resistance. $V_{oc}$ and $R_b$ are both considered as the quadratic functions of $SOC$.
\begin{align}\label{eq:eoc}
V_{oc} & =  b_{1}SOC^2 +b_{2}SOC+b_{3}\\
R_b & = c_{1}SOC^2 +c_{2}SOC +c_{3}
\end{align}

\noindent where $b_1$, $b_2$, $b_3$, $c_1$, $c_2$ and $c_3$ are coefficients.

\section{Optimization Problem Formulation}

The optimization problems are analyzed in MPC framework. The time step is set as 0.1s. Considering the calculation burden and control performance, the predictive horizon of MPC is set as 0.8s. In addition, we assume that the host vehicle can obtain the perfect information of the preceding vehicle's future velocity in every predictive horizon.

\subsection{Sequential Optimization}
\begin{figure}[!htb]
  \centering
  \includegraphics[width=\hsize]{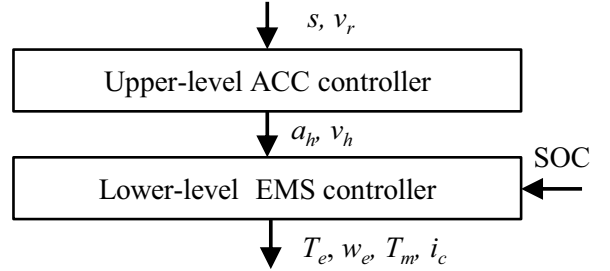}
  \caption{Sequential optimization structure}
  \label{fig2}
\end{figure}

The sequential optimization is used to evaluate co-optimization, which is shown in Fig.~\ref{fig2}. In this method, the upper-level ACC controller outputs the acceleration demand, which is given to lower-level EMS controller to optimize the acceleration demand, engine demand and the clutch engagement/disengagement demand. These two controllers are both based on MPC.

In ACC problem, the system states $ {x_{acc}}$ and control input $ {u_{acc}}$ are defined as follows.
\begin{equation}
	\begin{gathered}
	x_{acc}=[s,v_{r}]^\mathrm{T}\\
	u_{acc}=[a_h]
	\end{gathered}
\end{equation}

\noindent For the purpose of safety and comfortability, $a_h$ and $v_h$ is constrained in a range.
\begin{equation}
 \Phi_{acc} =
 \left(
	\begin{gathered}
	 a_{min} \leq a_h \leq a_{max}\\
	 v_{min} \leq v_h \leq v_{max}
	\end{gathered}
	\right)
\end{equation}

\noindent The ACC cost function is defined as
\begin{equation}
\begin{aligned}
{min}\enspace J_{acc}=\int_0^{t_{f}} \left (\frac{e}{e_{nmax}}\right)^2+\left(\frac{v_{r}}{v_{r,nmax}}\right)^2 \\ +\left(\frac{a_h}{a_{min}}\right)^2+\left(\frac{j_h}{j_{nmax}}\right)^2  \,\mathrm{d}t
\end{aligned}
\end{equation}
\noindent where $e_{nmax}$, $v_{r,nmax}$ and $j_{nmax}$ are the nominal maximum distance error, relative velocity and jerk, respectively. $a_{min}$ is the lower bound of acceleration.

In EMS problem, the system states $ {x_{ems}}$ and control input ${u_{ems}}$ are defined as follows.
\begin{equation}
	\begin{gathered}
	x_{ems}=[SOC]\\
	u_{ems}=[T_{e},w_{e},T_{m},i_c]^\mathrm{T}
	\end{gathered}
\end{equation}

\noindent The constraints are defined as follows.
\begin{equation}
  \Phi_{ems} =
  \left(
	\begin{gathered}
   w_e  =0 \cup w_{e,min}\leq w_e \leq w_{e,max} \\
   T_e  =0 \cup T_{e,min}\leq T_e \leq T_{e,max}\\
   w_{m,min}\leq  w_m\leq w_{m,max}\\
   T_{m,min}\leq  T_m\leq T_{m,max}\\
   SOC_{min} \leq SOC \leq SOC_{max}\\
   i_c \in \{0,1\}\\
	\end{gathered}
	\right)
\end{equation}

\noindent The EMS cost function is defined as
\begin{equation}
{min}\enspace J_{ems}=\int_0^{t_{f}} \dot m_{f}+\lambda \left( \frac{SOC-SOC_{ref}}{SOC_{max}-SOC_{min}}\right)^2   \,\mathrm{d}t
\end{equation}

\noindent where $SOC_{ref}$ is the constant reference of $SOC$, $\lambda$ is the cost coefficient, $SOC_{min}$ and $SOC_{max}$ are the minimum and maximum value of SOC. The nonlinear constrains and the integer variables constitute the MINLP.

\subsection{Co-optimization}

\begin{figure}[!htb]
  \centering
  \includegraphics[width=\hsize]{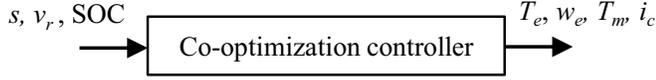}
  \caption{Co-optimization structure}
  \label{fig3}
\end{figure}

As shown in Fig.~\ref{fig3}, the co-optimization controller optimizes the ACC problem and EMS problem simultaneously. 

In this problem, the system states $ {x_{co}}$ and control input $ {u_{co}}$ are defined in Eqn.~29.
\begin{equation}
	\begin{gathered}
	x_{co}=[s,v_r,SOC]^\mathrm{T}\\
	u_{co}=[a_h,T_{e},w_{e},T_{m},i_c]^\mathrm{T}
	\end{gathered}
\end{equation}

The constraints are the combination of $\Phi_{acc}$ and $\Phi_{ems}$.  
\begin{equation}
	\Phi_{co}=[\Phi_{acc},\Phi_{ems}]^\mathrm{T}
\end{equation}

The cost function of co-optimization is defined as follows.
\begin{equation}
{min} \quad J_{co}=\int_0^{t_f} J_{acc}+J_{ems}  \,\mathrm{d}t
\end{equation}

Due to the increase of variables and constraints and the complexity of the cost function, co-optimization is a much larger and more complex MINLP than sequential optimization.


\section{Simulation}

In this section, the preceding vehicle follows the drive cycles. Worldwide harmonized light-duty vehicles test cycles (WLTC) is used in simulation. The initial value of SOC is set as 0.6, which assumes that the host vehicle is in forced charge sustaining mode. The optimization problems are solved by Gurobi 10.0 using branch-and-cut method. All the simulations are conducted on a desktop computer with a 12-core Intel i7 CPU and 32GB RAM.

\subsection{ ACC Performance Comparison}

\begin{figure}[!htb]
  \centering
  \includegraphics[width=\hsize]{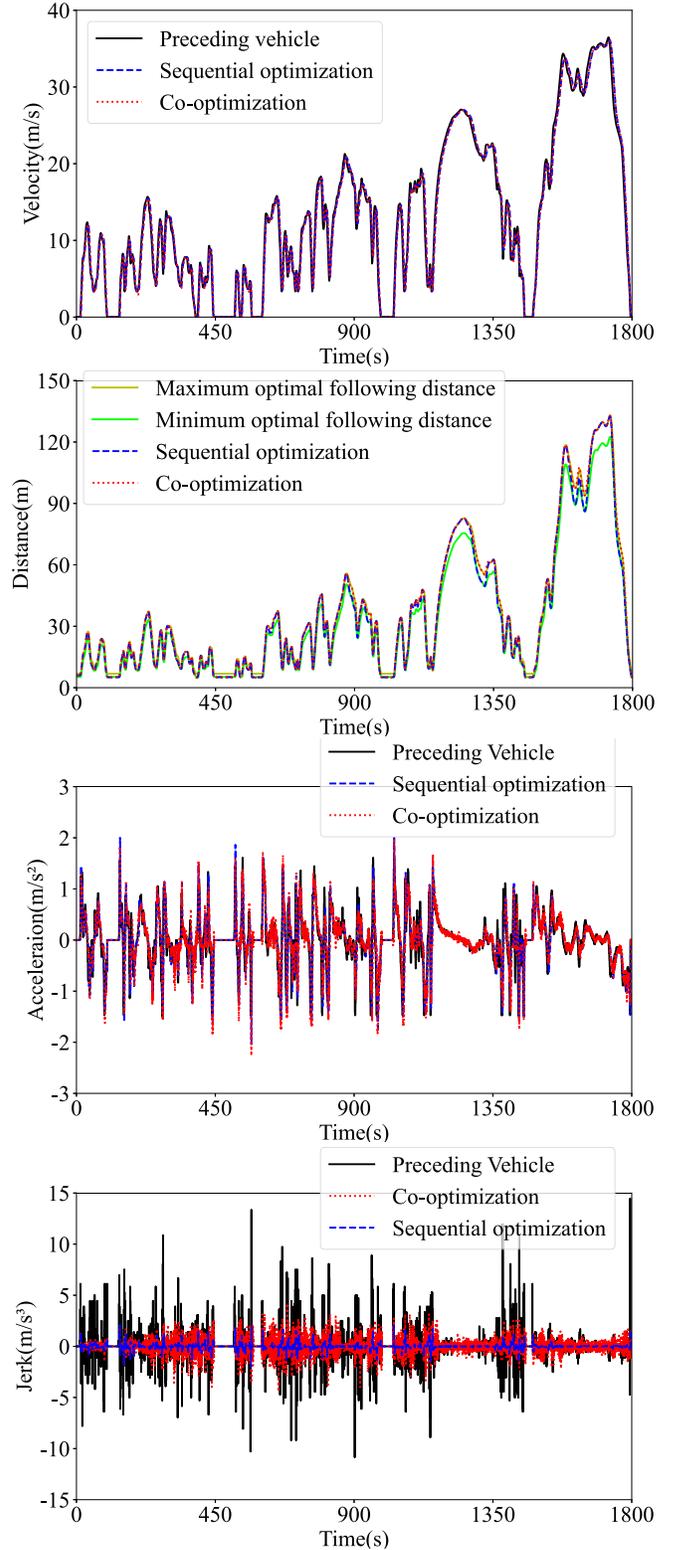}
  \caption{The ACC performance of sequential optimization and co-optimization}
  \label{fig4}
\end{figure}

Fig.~\ref{fig4} shows the ACC performance of sequential optimization and co-optimization. Both two methods can follow the velocity of the preceding vehicle and maintain the appropriate distance. In most cases, the acceleration and jerk of host vehicle are smoother than that of the preceding vehicle, which means that the host vehicle has the better driving comfort and energy saving. In sequential optimization, the jerks are in a range of $\pm±$ 2$m/s^3$, which is much less than those in co-optimization. The higher jerk will reduce the driving comfort, which is the tradeoff with fuel reduction.

\subsection{ EMS Performance Comparison}

\begin{figure}[!htb]
  \centering
  \includegraphics[width=\hsize]{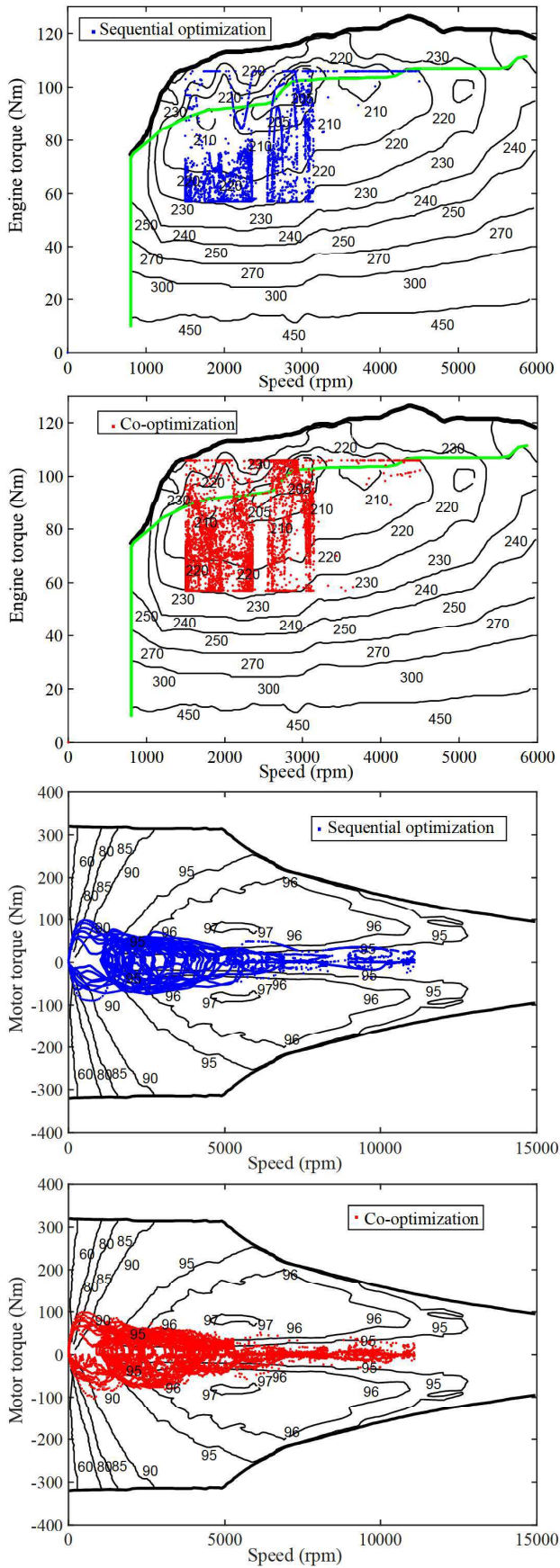}
  \caption{Operation points of engine and motor}
  \label{fig5}
\end{figure}

The operation points of engine and motor are exhibited in Fig.~\ref{fig5}. The green line is engine optimal operating points (OPP) line.

A majority of engine operation points can be optimized in high-efficiency area. Due to the engine torque constrains, most of the engine operation points distribute in the area where $T_e$ is larger than 57 Nm and $b_e$ is less than 230g/(kwh). Compare with sequential optimization, co-optimization distributes more operation points in the area of engine map with 210g/(kwh) and the area near OPP. For these two methods, most of the operation points of motor distribute in area of motor map with efficiency over than 90$\%$.

\begin{figure}[!htb]
  \centering
  \includegraphics[width=\hsize]{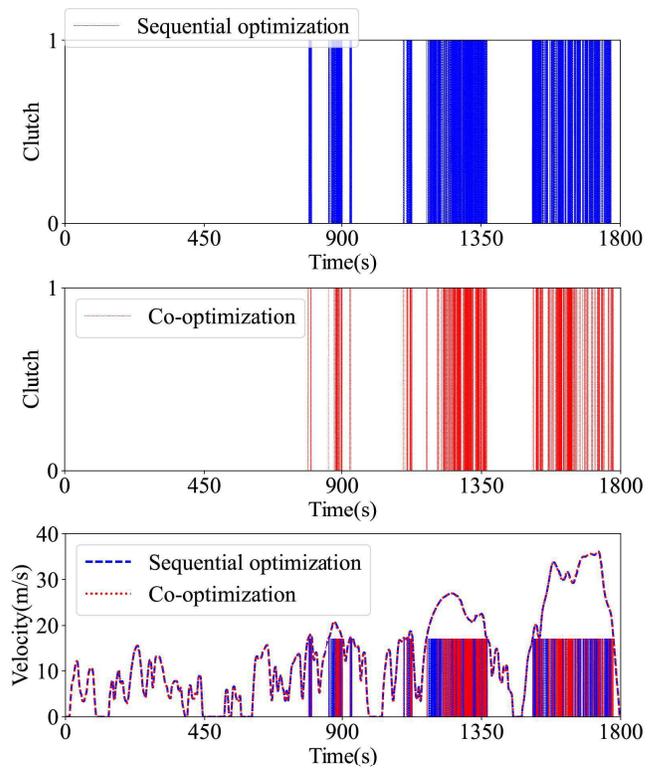}
  \caption{Clutch engagement/disengagement demand}
  \label{fig6}
\end{figure}

Fig.~\ref{fig6} shows the clutch engagement/disengagement demand. Based on co-optimization, the clutch engagement and disengagement frequencies are lower than that of sequential optimization. The clutch disengages at the velocity less than 17m/s, which signifies that the host vehicle is in the series mode. The engine drives the wheels directly when the host vehicle drives at high velocity with parallel mode.

\begin{figure}[!htb]
  \centering
  \includegraphics[width=\hsize]{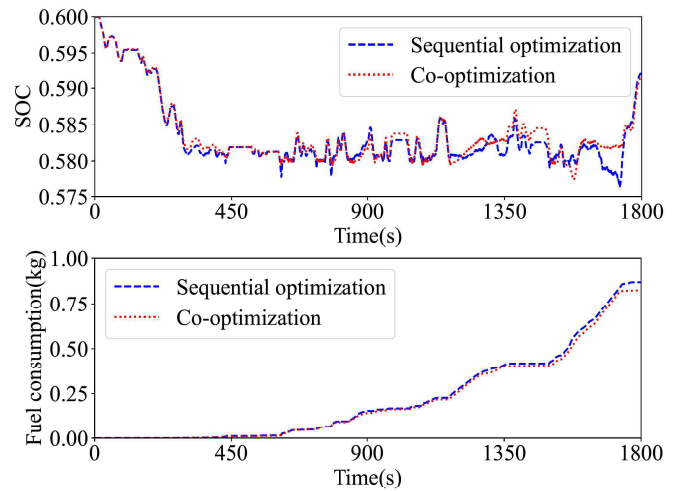}
  \caption{SOC and fuel consumption}
  \label{fig7}
\end{figure}

Fig.~\ref{fig7} shows the trend of SOC and fuel consumption. Co-optimization could always maintain lower fuel consumption. When the host vehicle drives at high velocity, the $SOC$ curve of these two methods are significantly different. It demonstrates the different energy distribution between these two optimization methods. The area where the fuel consumption is horizontal demonstrates that the host vehicle is driven in pure electric mode.

Table.~\ref{tab2} shows the comparisons of energy saving between these two methods. The final $SOC$ of co-optimization is 0.5914, which is only about 0.1$\%$ lower than sequential optimization. However, the fuel consumption of co-optimization is 5.1$\%$ less than sequential optimization. After converting the mass of fuel consumption to the consumption per hundred kilometers, the fuel consumption of the two methods are 3.736L/(100 km) and 3.545L/(100 km). In other words, co-optimization has better fuel economy than sequential optimization.

\begin{table}[!htb]
  \centering
  \caption{Comparison of fuel consumption and final SOC}
  \label{tab2}
  \begin{tabular}{c|c|c}
    \hhline    
    Method                  & Fuel consumption & Final SOC \\ \hline
    Sequential optimization &0.8692kg (0$\%$)             & 0.5920 (0$\%$)       \\ \hline
    Co-optimization        &0.8248kg (-5.1$\%$)        & 0.5914 (-0.1$\%$)       \\ \hline
    \hhline
  \end{tabular}
\end{table}

\subsection{Simulation Time Comparison}

Table.~\ref{tab3} shows the simulation time of these two methods. Due to the increase in numbers of state and control variables, it takes more time to compute the co-optimization problem, which is nearly 30 times more than that of sequential optimization. If co-optimization is applied to hardware-in-the-loop test or real vehicle experiment, it is dependent on high performance computers to ensure the real-time performance.

\begin{table}[!htb]
  \centering
  \caption{Simulation time}
  \label{tab3}
  \begin{tabular}{c|c}
    \hhline    
    Method                  & Average simulation time per time step \\ \hline
    Sequential optimization &  0.17s (0$\%$)                \\ \hline
    Co-optimization       &  5.26s (2994$\%$)                \\ \hline
    \hhline
  \end{tabular}
\end{table}

\section{Conclusion}
This paper presents the control-oriented EMS model of BYD Qin Plus DM-i and applies it to the co-optimization of ACC and EMS based on MPC framework. Though slightly less comfortable than sequential optimization, co-optimization shows excellent performance in EMS and saving 5.1$\%$ of fuel consumption compared to sequential optimization. However, the simulation time gap between co-optimization and sequential optimization is significant, which hinders the practical deployment. 

For the future work, learning-based method will be applied to co-optimization for the purpose of reducing online computation time. The results of this paper can also be used as a benchmark for the comparison of learning-based method to co-optimization.

\end{document}